 \documentclass[final,3p,times,twocolumn]{elsarticle}
\usepackage{amssymb}
\biboptions{sort&compress}

\journal{NIM B351(2015)6}

\begin{document}

\begin{frontmatter}



\title{New cross section data and review of production routes of medically used $^{110m}$In}


\author[1]{F. T\'ark\'anyi}
\author[1]{S. Tak\'acs}
\author[1]{F. Ditr\'oi\corref{*}}
\author[2]{A. Hermanne}
\author[3]{M. Baba}
\author[4]{B.M.A. Mohsena}
\author[5]{A.V. Ignatyuk}
\cortext[*]{Corresponding author: ditroi@atomki.hu}

\address[1]{Institute for Nuclear Research, Hungarian Academy of Sciences (ATOMKI),  Debrecen, Hungary}
\address[2]{Cyclotron Laboratory, Vrije Universiteit Brussel (VUB), Brussels, Belgium}
\address[3]{Cyclotron Radioisotope Center (CYRIC), Tohoku University, Sendai, Japan}
\address[4]{Nuclear Research Center – Egyptian Atomic Energy Authority, Cairo, Egypt}
\address[5]{Institute of Physics and Power Engineering (IPPE), Obninsk, Russia}

\begin{abstract}
Evaluation of nuclear data for production routes of $^{110m}$In is in progress in the frame of an IAEA Coordinated Research Project (CRP). New experimental cross section data for the indirect $^{nat}$In(p,x)$^{110}$Sn $\longrightarrow$ $^{110m}$In and for the direct $^{107}$Ag($\alpha$,n)$^{110m}$In and $^{109}$Ag($^{3}$He,2n)$^{110m}$In production  routes and for the satellite impurity reactions $^{107}$Ag($\alpha$,xn)$^{110g,109}$In and $^{109}$Ag($^{3}$He,xn)$^{110g,111,109}$In   have been measured by using the activation method, stacked foil irradiation technique and gamma-ray spectrometry. Additional data are reported for production of the $^{111}$In diagnostic gamma-emitter via the $^{109}$Ag($\alpha$,2n)$^{111}$In reaction. The earlier experimental data were critically reviewed in order to prepare recommended data and optimal production parameters for the different routes. 
\end{abstract}


\end{frontmatter}

% \linenumbers

\section{Introduction}
\label{1}
Various radiopharmaceuticals, labeled with $^{111}$In (T$_{1/2}$ = 2.81 d, 100 \% EC decay) are used in the diagnosis of cancer and other diseases through SPECT. However, in particular for receptor-type studies, the quantification of the uptake of the radiopharmaceutical via PET measurements might be important. For $^{111}$In the corresponding isotope of choice is $^{110m}$In. This metastable state with I$^{\pi}$ = $2^+$ has a half-life of 69.1 min and decays in 61.25 \% via positron emission with E$_{max}$ = 2.3 MeV. Evaluation of nuclear data for production routes of $^{110m}$In is performed in the frame of an IAEA Coordinated Research Project \cite{1}.  The task was assigned to the ATOMKI-VUB group, which previously already published results for some related reactions \cite{2,3,4,5,6,7,8,9}.  As the compilation of all earlier experimental studies showed a lack of data and in some cases large disagreements, we decided to re-investigate the most promising reactions.
Two basic routes exist for production of $^{110m}$In:  a direct and an indirect process. The principal reactions of direct production, with limited contaminants, are $^{110}$Cd(p,n), $^{110}$Cd(d,2n),  $^{107}$Ag($\alpha$,n) and $^{109}$Ag($^{3}$He,2n). The direct production of $^{110m}$In always leads to the co-formation of the ground state $^{110g}$In (I$^{\pi}$= $7^+$, T$_{1/2}$ = 4.9 h). This co-formation increases with increasing projectile energy because of the higher spin of the ground state, but it cannot be observed in all cases of our investigations.  In all cases, in order to minimize the impurity level, highly enriched targets should be used and the covered energy range should be optimized, which requires information on the production of the disturbing, simultaneously produced, longer-lived $^{109}$In, $^{110g}$In and $^{111}$In radioisotopes.
Isotopically pure $^{110m}$In could only be prepared via the generator system $^{110}$Sn (T$_{1/2}$ = 4.11 h, EC 100 \%)$\longrightarrow$ $^{110m}$In. The mother isotope of the generator ($^{110}$Sn) could be obtained at low and medium energies with light charged particles via the $^{113}$In(p,4n), $^{113}$In(d,5n), $^{110}$Cd($^{3}$He,3n), $^{108}$Cd($\alpha$,2n) and $^{110}$Cd($\alpha$,4n) reactions. Disturbing co-produced radioisotopes, requiring optimization of the energy range and use of enriched targets, are the long-lived $^{113g}$Sn, decaying to $^{113m}$In and finally to stable $^{113}$In (hence decreasing specific activity), and the rather short-lived $^{111}$Sn (T$_{1/2}$ = 35.3 min) decaying to $^{111}$In (T$_{1/2}$= 2.81 d).

\section{Experimental}
\label{2}
The general characteristics and procedures for irradiation, activity assessment and data evaluation (including estimation of uncertainties) were similar to many of our earlier works \cite{10, 11}. The main experimental parameters and the methods of data evaluation for the present study are summarized in Table 1. The used decay data are collected in Table 2.

\begin{table*}[t]
\tiny
\caption{Main experimental parameters and methods of data evaluation}
\centering
\begin{center}
\begin{tabular}{|p{0.7in}|p{0.8in}|p{0.8in}|p{0.8in}|p{0.8in}|p{0.8in}|p{0.8in}|}
\hline
\multicolumn{5}{|c|}{\textbf{Experiment}} & \multicolumn{2}{|c|}{\textbf{Data evaluation}} \\
\hline
Reaction & $^{nat}$In(p,x) & $^{nat}$In(p,x) & $^{nat}$Ag($\alpha$,x) 
& $^{nat}$Ag($^{3}$He,x) & & \\
\hline
Incident particle & Proton & Proton & $\alpha$-particle & $^{3}$He & Gamma 
spectra evaluation & Genie 2000, Forgamma \cite{12, 13} \\
\hline
Method & Stacked foil & Stacked foil & Stacked foil & Stacked foil & 
Determination of beam intensity & Faraday cup (preliminary)Fitted 
monitor reaction (final) \cite{14} \\
\hline
Target stack and thicknesses & Er(25 $\mu$m) Co(50 $\mu$m) Al(10 $\mu$m) In(50 
$\mu$m) Al(100 $\mu$m) block\newline Repeated 13 times & Al(11 $\mu$m) In(116.3 $\mu$m) Al(99.2 
$\mu$m) V(8.41 $\mu$m) Al(99.2 $\mu$m) Ho(26.2 $\mu$m) Al(99.2 $\mu$m) block\newline Repeated 19 times 
& 2 stacks, each 8 Ag(8.1 $\mu$m) 6 Cu(7.4 $\mu$m) & 17 Ag(8.1 $\mu$m) 3 Ti(12 $\mu$m) 
& Decay data & NUDAT 2.6 \cite{15} \\
\hline
Number of target foils & 13 & 19 & 16 & 17 & Reaction Q-values & Q-value 
calculator \cite{16} \\
\hline
Accelerator & K110 MeV cyclotron CYRIC, Sendai & K110 MeV cyclotron LLN 
Louvain-la-Neuve & CGR-560 cyclotron VUB Brussels & MGC -20 
cyclotron ATOMKI Debrecen & Determination of beam energy & Andersen and 
Ziegler (preliminary)\cite{17} Fitted monitor reaction (final) \cite{17} \\
\hline
Primary energy & 70 MeV & 65 MeV & 27 MeV & 26 MeV & Uncertainty of 
energy & Cumulative effects of possible uncertainties \\
\hline
Irradiation time & 61 min & 60 min & 60 min & 47 min & Cross sections & 
Isotopic cross section \\
\hline
Beam current & 52 nA & 50 nA & 120 nA & 101 nA & Uncertainty of cross 
sections & Sum in quadrature of all individual contribution \cite{18} \\
\hline
Monitor reaction, $[$recommended values$]$ & $^{27}$Al(p,x)$^{22,24
}$Na reaction \cite{19} & $^{27}$Al(p,x)$^{22,24}$Na reaction 
$[$19$]$ & $^{nat}$Cu ($\alpha$,x)$^{66,67}$Ga reaction \cite{19} & $^{
nat}$Ti($^{3}$He,x)$^{48}$V reaction $[$19$]$ & Yield & 
Physical yield \cite{20} \\
\hline
Monitor target and thickness & $^{27}$Al, 100 $\mu$m & $^{27}$Al, 
99.2 $\mu$m & $^{nat}$Cu, 7.4 $\mu$m & $^{nat}$Ti, 12 ?m & Theory & 
ALICE-IPPE \cite{21},EMPIRE \cite{22},TALYS (TENDL 2013, 2014) \cite{23} \\
\hline
detector & HPGe & HPGe & HPGe & HPGe & & \\
\hline
$\gamma$-spectra measurements & 2 series & 2 series & 4 series & 2 series & & 
\\
\hline
Cooling times & 22-30 h\newline 315-340 h & 8-12 h\newline 141-193 h & 0.3-2.5 h\newline 5.1-8.3 
h \newline 22.1-48.9 h\newline 52-99.7 h & 0.5-3.7 h\newline 3.9-11.8 & & \\
\hline
\end{tabular}

\end{center}
\end{table*}

\begin{table*}[t]
\tiny
\caption{Decay characteristics of the investigated reaction products and Q-values of reactions for their productions}
\centering
\begin{center}
\begin{tabular}{|p{0.7in}|p{0.5in}|p{0.5in}|p{0.5in}|p{0.7in}|p{0.7in}|}
\hline
\textbf{Nuclide\newline Decay mode\newline Level energy} & \textbf{Half-life} & \textbf{E$_{\gamma}$(keV)} & \textbf{I$_{\gamma 
}$(\%)} & \textbf{Contributing reaction} & \textbf{Q-value(keV) }\\
\hline
$^{109}$In\newline $\beta^{+}$: 6.6 \% EC: 93.4 \% & 4.167 h & 
331.2\newline 649.8\newline 1099.2\newline 1321.3 & 9.7\newline 28\newline 30\newline 11.9 & $^{107}$Ag($^{3}$He,n)\newline 
$^{109}$Ag($^{3}$He,3n)\newline $^{107}$Ag($\alpha$,2n)\newline $^{109}$Ag($\alpha$,4n)\newline 
$^{110}$Cd(p,2n)\newline $^{111}$Cd(p,3n)\newline $^{112}$Cd(p,4n)\newline $^{113}$
Cd(p,5n)\newline $^{114}$Cd(p,6n)\newline $^{116}$Cd(p,8n)\newline $^{108}$Cd(d,n)\newline $^{
110}$Cd(d,3n)\newline $^{111}$Cd(d,4n)\newline $^{112}$Cd(d,5n)\newline $^{113}$
Cd(d,6n)\newline $^{114}$Cd(d,7n)\newline $^{116}$Cd(d,9n) & 
4941.24\newline -11514.7\newline -15636.39\newline -32092.32\newline -12714.5\newline -19690.14\newline -29084.19\newline -35622.95\newline -44665.87\newline -59506.21\newline 2299.78\newline -14939.08\newline -21914.71\newline -31308.76\newline -37847.52\newline -46890.43\newline -61730.77 
\\
\hline
$^{110m}$In\newline $\beta^{+}$ : 61.25 \%EC: 38.75 \%\newline 62.084 keV & 
69.1 min &  657.75\newline  818.05\newline 1125.77 & 97.74\newline 0.87\newline  1.04 & $^{109}$Ag($^{3
}$He,2n)\newline $^{107}$Ag($\alpha$,n)\newline $^{109}$Ag($\alpha$,3n)\newline $^{110}$Cd(p,n)\newline $^{
111}$Cd(p,2n)\newline $^{112}$Cd(p,3n)\newline $^{113}$Cd(p,4n)\newline $^{114}$
Cd(p,5n)\newline $^{116}$Cd(p,7n)\newline $^{110}$Cd(d,2n)\newline $^{111}$Cd(d,3n)\newline $^{
112}$Cd(d,4n)\newline $^{113}$Cd(d,5n)\newline $^{114}$Cd(d,6n)\newline $^{116}$
Cd(d,8n) & 
-3460.5\newline -7582.2\newline -24038.2\newline -4660.3\newline -11636.0\newline -21030.0\newline -27568.8\newline -36611.7\newline -51452.1\newline -6884.9\newline -13860.6\newline -23254.6\newline -29793.4\newline -38836.3\newline -53676.6 
\\
\hline
$^{110g}$In\newline $\beta^{+}$ : 0.008 \%EC: 99.992 \% & 4.92 h & 
641.68\newline 657.750\newline 707.40\newline 884.667\newline 937.478\newline 997.16 &  26.098\newline  29.5\newline 93\newline 68.4\newline 10.5 & $^{
109}$Ag($^{3}$He,2n)\newline $^{107}$Ag($\alpha$,n)\newline $^{109}$Ag($\alpha$,3n)\newline $^{
110}$Cd(p,n)\newline $^{111}$Cd(p,2n)\newline $^{112}$Cd(p,3n)\newline $^{113}$
Cd(p,4n)\newline $^{114}$Cd(p,5n)\newline $^{116}$Cd(p,7n)\newline $^{110}$Cd(d,2n)\newline $^{
111}$Cd(d,3n)\newline $^{112}$Cd(d,4n)\newline $^{113}$Cd(d,5n)\newline $^{114}$
Cd(d,6n)\newline $^{116}$Cd(d,8n) & 
-3460.5\newline -7582.2\newline -24038.2\newline -4660.3\newline -11636.0\newline -21030.0\newline -27568.8\newline -36611.7\newline -51452.1\newline -6884.9\newline -13860.6\newline -23254.6\newline -29793.4\newline -38836.3\newline -53676.6 
\\
\hline
$^{111}$In\newline EC: 100 \% & 2.81 d &  171.28\newline 245.35 & 90.7\newline 94.1  & $^{109
}$Ag($^{3}$He,n)\newline $^{109}$Ag($\alpha$,2n)\newline $^{111}$Cd(p,n)\newline $^{112}$
Cd(p,2n)\newline $^{113}$Cd(p,3n)\newline $^{114}$Cd(p,4n)\newline $^{116}$Cd(p,6n)\newline $^{
110}$Cd(d,n)\newline $^{111}$Cd(d,2n)\newline $^{112}$Cd(d,3n)\newline $^{113}$
Cd(d,4n)\newline $^{114}$Cd(d,5n)\newline $^{116}$Cd(d,7n) & 
6530.92\newline -14046.71\newline -19690.14\newline -11038.57\newline -17577.34\newline -26620.26\newline -41460.61\newline 3106.54\newline -3869.1\newline -13263.15\newline -19801.91\newline -28844.83\newline -43685.18 
\\
\hline
$^{109}$Sn\newline $\beta^{+}$:6.6 \% EC: 83.4 \%  & 18.0 min & 
649.8\newline 1099.2\newline 1321.3 & 28\newline  30\newline 11.9  & $^{113}$In(p.5n)$^{115}$In(p,7n)\newline 
$^{108}$Cd($^{3}$He,2n)\newline $^{110}$Cd($^{3}$He,4n)\newline $^{111}$
Cd($^{3}$He,5n)\newline $^{112}$Cd($^{3}$He,6n)\newline $^{113}$Cd($^{3
}$He,7n)\newline $^{114}$Cd($^{3}$He,8n)\newline $^{106}$Cd($\alpha$,n)\newline $^{108}$
Cd($\alpha$,3n)\newline $^{110}$Cd($\alpha$,5n)\newline $^{111}$Cd($\alpha$,6n)\newline $^{112}$Cd($\alpha$,7n)\newline $^{
113}$Cd($\alpha$,8n)\newline $^{114}$Cd($\alpha$,10n) & 
-39802.54\newline -56115.7\newline -7833.06\newline -25071.91\newline -32047.55\newline -41441.59\newline -47980.36\newline -57023.26\newline -10147.5\newline -28410.69\newline -45649.54\newline -52625.16\newline -62019.21\newline -68557.96 
\\
\hline
\end{tabular}
\end{center}
\end{table*} 

\setcounter{table}{1}

\begin{table*}[t]
\tiny
\caption{continued}
\centering
\begin{center}
\begin{tabular}{|p{0.7in}|p{0.5in}|p{0.5in}|p{0.5in}|p{0.7in}|p{0.7in}|}
\hline

$^{110}$Sn\newline EC: 100 \%  & 4.167 h & 280.462 & 97 & $^{108}$Cd($^{3
}$He,n)\newline $^{110}$Cd($^{3}$He,3n)\newline $^{111}$Cd($^{3}$He,4n)\newline 
$^{112}$Cd($^{3}$He,5n)\newline $^{113}$Cd($^{3}$He,6n)\newline $^{114}$
Cd($^{3}$He,7n)\newline $^{116}$Cd($^{3}$He,9n)\newline $^{108}$Cd($\alpha$,2n)\newline 
$^{110}$Cd($\alpha$,4n)\newline $^{111}$Cd($\alpha$,5n)\newline $^{112}$Cd($\alpha$,6n)\newline $^{113}$
Cd($\alpha$,7n)\newline $^{114}$Cd($\alpha$,8n)\newline $^{113}$In(p,4n)\newline $^{115}$In(p,6n) & 
3449.3\newline -13789.5\newline -20765.2\newline -30159.2\newline -36698.0\newline -45740.9\newline -60581.2\newline -17128.3\newline -34367.1\newline -41342.8\newline -50736.8\newline -57275.6\newline -66318.5\newline -28520.1\newline -44833.3 
\\
\hline
$^{111}$Sn\newline $\beta^{+}$ : 30.2009 \% EC: 69.7991 \% & 35.3 min & 
761.971152.98 & 1.482.7 & $^{110}$Cd($^{3}$He,2n)\newline $^{111}$Cd(
$^{3}$He,3n)\newline $^{112}$Cd($^{3}$He,4n)\newline $^{113}$Cd($^{3}$
He,5n)\newline $^{114}$Cd($^{3}$He,6n)\newline $^{116}$Cd($^{3}$He,8n)\newline $^{
108}$Cd($\alpha$,n)\newline $^{110}$Cd($\alpha$,3n)\newline $^{111}$Cd($\alpha$,4n)\newline $^{112}$
Cd($\alpha$,5n)\newline $^{113}$Cd($\alpha$,6n)\newline $^{114}$Cd($\alpha$,7n)\newline $^{116}$Cd($\alpha$,9n)\newline $^{
113}$In(p,3n)\newline $^{115}$In(p,5n) & 
-5620.67\newline -12596.3\newline -21990.35\newline -37572.03\newline -37572.03\newline -52412.37\newline -8959.43\newline -26198.3\newline -33173.93\newline -42567.98\newline -49106.73\newline -58149.64\newline -72989.98\newline -20351.3\newline -36664.46 
\\
\hline
\end{tabular}

\end{center}
\end{table*}

\section{Experimental results}
\label{3}
We are reporting cross sections for production of $^{110m}$In, $^{109}$In, $^{110g}$In and $^{111}$In in Ag targets irradiated with $^{3}$He and alpha particles.  The last three radioisotopes are important from the point of view of radionuclide purity in the case of direct production of $^{110m}$In. For the experiments on In targets, proton induced cross section data for production of the mother isotope $^{110}$Sn are reported and data for the shorter-lived, co-produced $^{109}$Sn and $^{111}$Sn (causing radionuclide impurity through decay to $^{109}$In and $^{111}$In) are discussed.
The numerical data of the measured cross sections are collected in Table 3-5. They are shown in graphical form in the corresponding sections discussing the different production routes.

\section{Production routes}
\label{4}

\subsection{Indirect}
\label{4.1}

The medically useful $^{110m}$In is produced through the 100 \% EC decay of $^{110}$Sn. The production routes for $^{110}$Sn include proton induced reactions at low and medium energy on stable isotopes of indium or alpha-particle and $^{3}$He induced reactions on stable isotopes of cadmium. 

\subsubsection{In+p}
\label{4.1.1}

The mother isotope $^{110}$Sn can be obtained on the two stable isotopes of indium ($^{113}$In: 4.3 \% and $^{115}$In: 95.7 \%), through the $^{113}$In(p,4n) and $^{115}$In(p,6n) reactions. As the high mass target isotope has an abundance of more than 95\%, the (p,6n) reaction, needing incident energies of around 60 MeV, should be favorite for efficient production. From the point of view of radio-purity in any case, shorter-lived $^{111}$Sn (decays to contaminating $^{111}$In) and $^{113}$Sn (decays to contaminating $^{113m}$In) will be produced that will also have an influence on the specific activity of the final product. Presence of $^{111}$Sn can be minimized by choosing the used energy range, irradiation time, or/and a proper cooling time before loading of the generator to let shorter-lived $^{111}$Sn (and lower mass short-lived Sn radioisotopes) decay. Some contamination with $^{113m}$Sn $\longrightarrow$ $^{113g}$Sn can never be avoided but as high energy protons are needed for the (p,4n) (threshold 28.8 MeV)  and (p,6n) (threshold 44.8 MeV) reactions, limited target thickness will insure that the cross section for $^{113}$In(p,n) in the target is low, as (p,n) reactions usually have maximum at much lower energies, so the protons leave the “thin” target before decelerating down into the energy range favorable for (p,n) reaction. Moreover, even if produced, the long half-life of $^{113g}$Sn results in low activity of its decay product $^{113m}$In and limited influence on the radio-purity. Radionuclides of Sn with mass lower than 110, and decaying to In radio-products, have even shorter half-life. Their presence could only be eliminated by keeping the incident proton energy below the threshold of the $^{113}$In(p,5n) (threshold 40.2 MeV) reaction at the cost of large production losses. No experimental cross section data were found for activation cross sections on monoisotopic targets, only data on $^{nat}$In were published. All available experimental activation cross section data for production of $^{109}$Sn (T$_{1/2}$ = 18 min), $^{110}$Sn (T$_{1/2}$ = 4.167 h) and $^{111}$Sn (T$_{1/2}$ = 35.3 min) are shown in Fig. 1. In our experiments we could deduce cross section data only for production of $^{110}$Sn due to the long cooling time after EOB (end of bombardment). Two earlier data sets exist, measured by Lundquist et al. \cite{24} and Nortier et al. \cite{25, 26}.  Our data for $^{110}$Sn are in acceptable agreement with those.
According to Fig. 1, no production of $^{109}$Sn is observable up to 55 MeV, but the amount of $^{111}$Sn is significant, resulting in high yield of the $^{111}$In decay product. At energies above 60 MeV (the threshold of the $^{115}$In(p,6n)$^{110}$Sn reaction is 44.8 MeV) the $^{111}$Sn/$^{110}$Sn ratio is becoming better, therefore this energy range is preferred, resulting in high yields of $^{110m}$In using targets with natural composition. The radionuclide purity can be improved significantly by using irradiation times up to 3 half-lives of $^{110}$Sn and a longer cooling time to let the simultaneously produced shorter-lived $^{109}$Sn and $^{111}$Sn decay.

\begin{figure}
\includegraphics[scale=0.3]{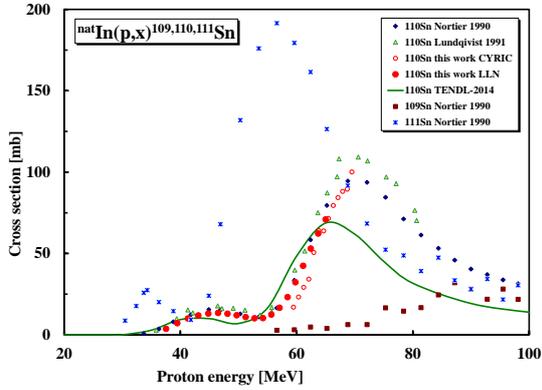}
\caption{Activation cross sections of proton induced nuclear reactions for $^{nat}$In(p,xn)$^{109,110,111}$Sn}
\end{figure}

\subsubsection{Cd+$\alpha$}
\label{4.1.2}

Another suggested route for the production of $^{110m}$In through the $^{110}$Sn/$^{110m}$In isotope generator is the $^{108}$Cd(α,2n)$^{110}$Sn nuclear reaction \cite{27} and at higher energies the $^{110}$Cd($\alpha$,4n) $^{110}$Sn reaction. We found only one relevant experimental data set, the production cross section of $^{110}$Sn on $^{nat}$Cd, published by us \cite{9}. No data for $^{109}$Sn and $^{111}$Sn were given. The comparison of these experimental values and the theoretical data from TENDL-2014 is shown in Fig. 2. The TENDL-2014 prediction underestimates the magnitude but the shape is well reproduced. As $^{nat}$Cd contains 8 stable isotopes ($^{106}$Cd -1.25 \%, $^{108}$Cd -0.89\%, $^{110}$Cd -12.49\%, $^{111}$Cd -12.80 \%, $^{112}$Cd -24.13 \%, $^{113}$Cd -12.22 \%, $^{114}$Cd -28.73 \%, $^{116}$Cd -7.49 \%) many reactions of the ($\alpha$,xn) type can contribute. When using $^{nat}$Cd the contribution through the $^{106}$Cd($\alpha, \gamma$) reaction is very small. The lowest number of contaminating radioisotopes is obtained by relying on the $^{108}$Cd($\alpha$,2n)$^{110}$Sn  reaction as only shorter-lived $^{111}$Sn will unavoidably be produced through the ($\alpha$,n) reaction. Choosing an appropriate cooling time, allowing the decay of $^{111}$Sn before preparing the generator, will result in practically nca (no carrier added) $^{110m}$In. In principle we can also use $^{nat}$Cd targets, because the additionally produced Sn isotopes are either stable or long-lived ($^{113}$Sn) and decay of this last results in negligible influence on the specific activity as discussed in the previous section. The highest yield and lowest contamination will be obtained by using highly enriched $^{108}$Cd. No experimental data were presented on $^{108}$Cd, but cross sections up to the threshold of $^{110}$Cd($\alpha$,4n)$^{110}$Sn reaction (35.6 MeV) can be deduced from  cross sections measured on $^{nat}$Cd. The theoretical cross sections for production of $^{109,110,111}$Sn by alpha irradiation of $^{nat}$Cd are shown in Fig. 2. It can be seen on Fig. 3 and from Table 2 that using highly enriched $^{108}$Cd targets, up to 30 MeV no contamination with $^{109}$Sn exists. By using the 24-30 MeV energy range and the above mentioned long irradiation and longer cooling time also the amount of $^{111}$Sn can be minimized. 

\begin{figure}
\includegraphics[scale=0.3]{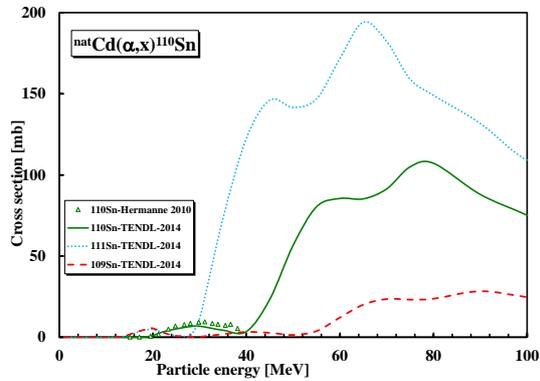}
\caption{Excitation functions of the  $^{nat}$Cd($\alpha$,xn)$^{109,110,111}$Sn  reactions}
\end{figure}

\begin{figure}
\includegraphics[scale=0.3]{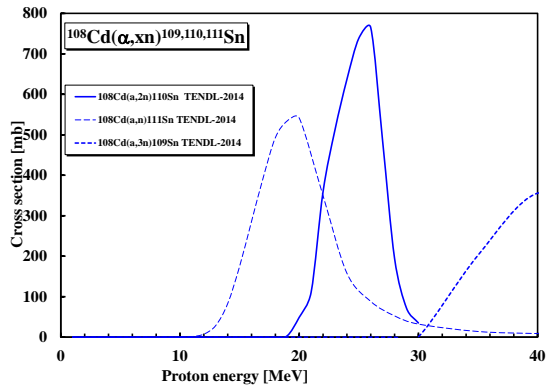}
\caption{Theoretical excitation functions of the  $^{108}$Cd($\alpha$,xn)$^{109,110,111}$Sn  reactions}
\end{figure}

\subsubsection{Cd+$^{3}$He}
\label{4.1.3}

The situation is nearly the same as for the Cd + $\alpha$ route. The radionuclide of interest, $^{110}$Sn, can also be produced through $^{nat}$Cd($^{3}$He,xn) or, with higher yield, using enriched targets through the  $^{110}$Cd($^{3}$He,3n) reaction. No experimental data exist for monoisotopic targets. Two earlier studies were published for production of $^{110}$Sn and $^{111}$Sn on $^{nat}$Cd \cite{2, 5} (reported by our group). The experimental data are shown in Fig. 4 in comparison with the TENDL-2014 calculation. A good agreement is seen between the two experimental datasets but the predictions of TENDL-2014 differ drastically, both in shape and in magnitude. 
The theoretical excitation functions of the $^{110}$Cd($^{3}$He,xn)$^{109,110,111}$Sn  reactions are shown in Fig. 5, but they are probably unrealistic as discussed above (according to the systematics on the neighboring elements the maximum cross sections  of the $^{110}$Cd ($^{3}$He,3n)  reaction should be around 500-600 mb). When comparing the $\alpha$ and $^{3}$He routes in the low energy region (up to 30 MeV) for the $^{108}$Cd($\alpha$,2n) and the $^{110}$Cd($^{3}$He,3n) reactions, it can be seen that the cross sections on $^{108}$Cd and $^{110}$Cd are similar for the corresponding particles, therefore due to the lower stopping, the yield of the $^{3}$He route should be higher. It is, however, well known that high intensity $^{3}$He beams are rare and very expensive, even when a gas recovery system is available at the accelerator.

\begin{figure}
\includegraphics[scale=0.3]{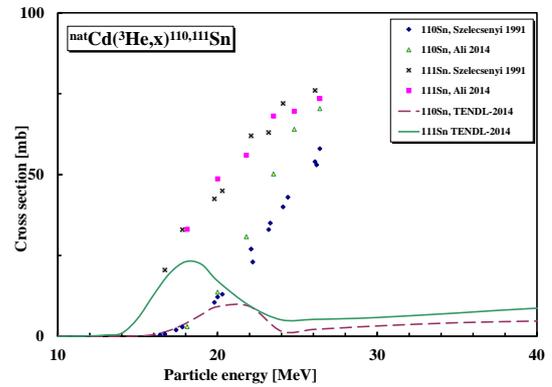}
\caption{Excitation functions of the  $^{nat}$Cd( $^{3}$He,xn)$^{110,111}$Sn  reactions}
\end{figure}

\begin{figure}
\includegraphics[scale=0.3]{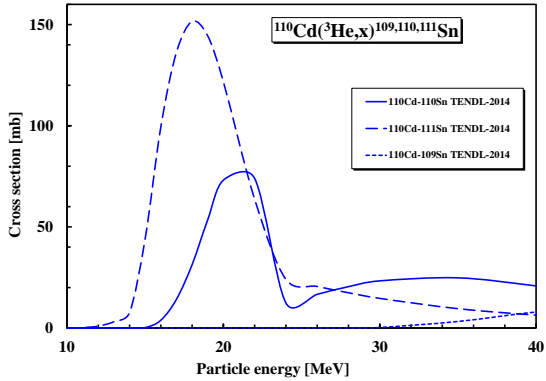}
\caption{Theoretical excitation functions of the  $^{110}$Cd($^{3}$He,xn)$^{109,110,111}$Sn  reactions}
\end{figure}

\section{Direct production}
\label{5}
The direct production routes include proton or deuteron induced reactions on cadmium and -particle or $^{3}$He induced reactions on silver.

\subsection{Cd+p}
\label{5.1}
Production of $^{110m}$In (T$_{1/2}$= 69.1 min), with not too high contamination, via proton and deuteron induced reactions on Cd can only be done on highly enriched targets. When $^{nat}$Cd is used, the many stable target isotopes lead to possible parallel production of various In radioisotopes with comparable or longer half-lives than $^{110m}$In: $^{108m}$In (T$_{1/2}$ = 39.6 min), $^{108g}$In (T$_{1/2}$ = 58 min), $^{109}$In (T$_{1/2}$ = 4.167 h), $^{111}$In (T$_{1/2}$ = 2.81 d), $^{113m}$In (T$_{1/2}$ = 99.48 min), $^{114m}$In (T$_{1/2}$ = 49.51 d), $^{115m}$In (T$_{1/2}$ = 4.436 h),  $^{116m}$In (T$_{1/2}$ = 54.29 min), $^{117}$In (T$_{1/2}$ = 43.2 min), $^{117m}$In (T$_{1/2}$ = 1.937 h).
By irradiating highly enriched $^{110}$Cd with protons in a limited energy range, only $^{110}$In is produced through a (p,n) reaction. As it was mentioned before in a direct production route not only the metastable state $^{110m}$In is produced, but also its longer-lived, higher spin $^{110g}$In ground state  (T$_{1/2}$= 4.92 h).
The experimental and theoretical excitation functions for the $^{110}$Cd(p,xn)110m,$^{110g}$In  reactions are shown in Fig. 6 and 7. For $^{109}$In production we present only the theoretical data in Fig. 7 to see the reaction threshold and the shape of the excitation function.
Experimental data for the cross section of $^{110}$Cd(p,n)$^{110m}$In reactions (Fig. 6) are available from Blaser et al. 1951 \cite{28}, Al Saleh et al. \cite{29}, Khandaker et al.  \cite{30}, Tárkányi et al. \cite{3}, Skakun et al. \cite{31}, Otozai et al. \cite{32}, Nortier et al. \cite{25} and Kormali  et al. \cite{33}. 
The data measured on $^{nat}$Cd were normalized for $^{110}$Cd target up to the threshold of the $^{111}$Cd(p,2n)$^{110}$In reaction.
In Fig. 7 the excitation function of the $^{110}$Cd(p,n)$^{110g}$In  reaction is presented. Experimental data on $^{nat}$Cd and $^{110}$Cd are available from Skakun et al. \cite{31}, Abramovich et al. \cite{34},  Otozai et al. \cite{32},  Tarkanyi et al. \cite{3} and Elbinawi et al. \cite{35}.  Fig. 7 shows that by properly selecting the incident energy, the contamination with $^{109}$In can be minimized. Comparison of the Fig. 6 and Fig. 7 shows that the useful energy ranges for $^{110m}$In and $^{110g}$In production are the same. The cross sections for $^{110g}$In are a factor of 6 higher and, in spite of the four times longer half-life, the activity ratio at EOB will still be unacceptable from the point of view of radionuclide purity. The $^{110m}$In and $^{110g}$In decay independently and the activity ratio during the irradiation will change in the target as the meta-state reaches saturation faster. Hence only rather short irradiations could help to reduce somewhat the contamination level. 

\begin{figure}
\includegraphics[scale=0.3]{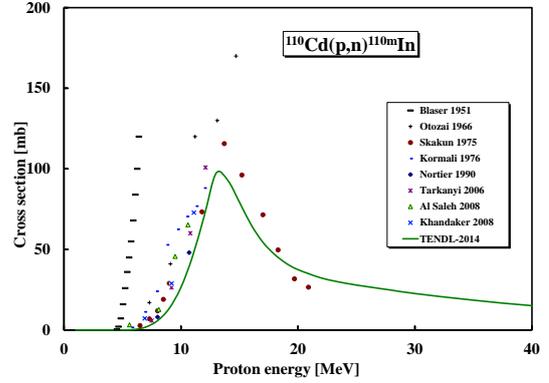}
\caption{Excitation functions of the $^{110}$Cd(p,n)$^{110m}$In reaction}
\end{figure}

\begin{figure}
\includegraphics[scale=0.3]{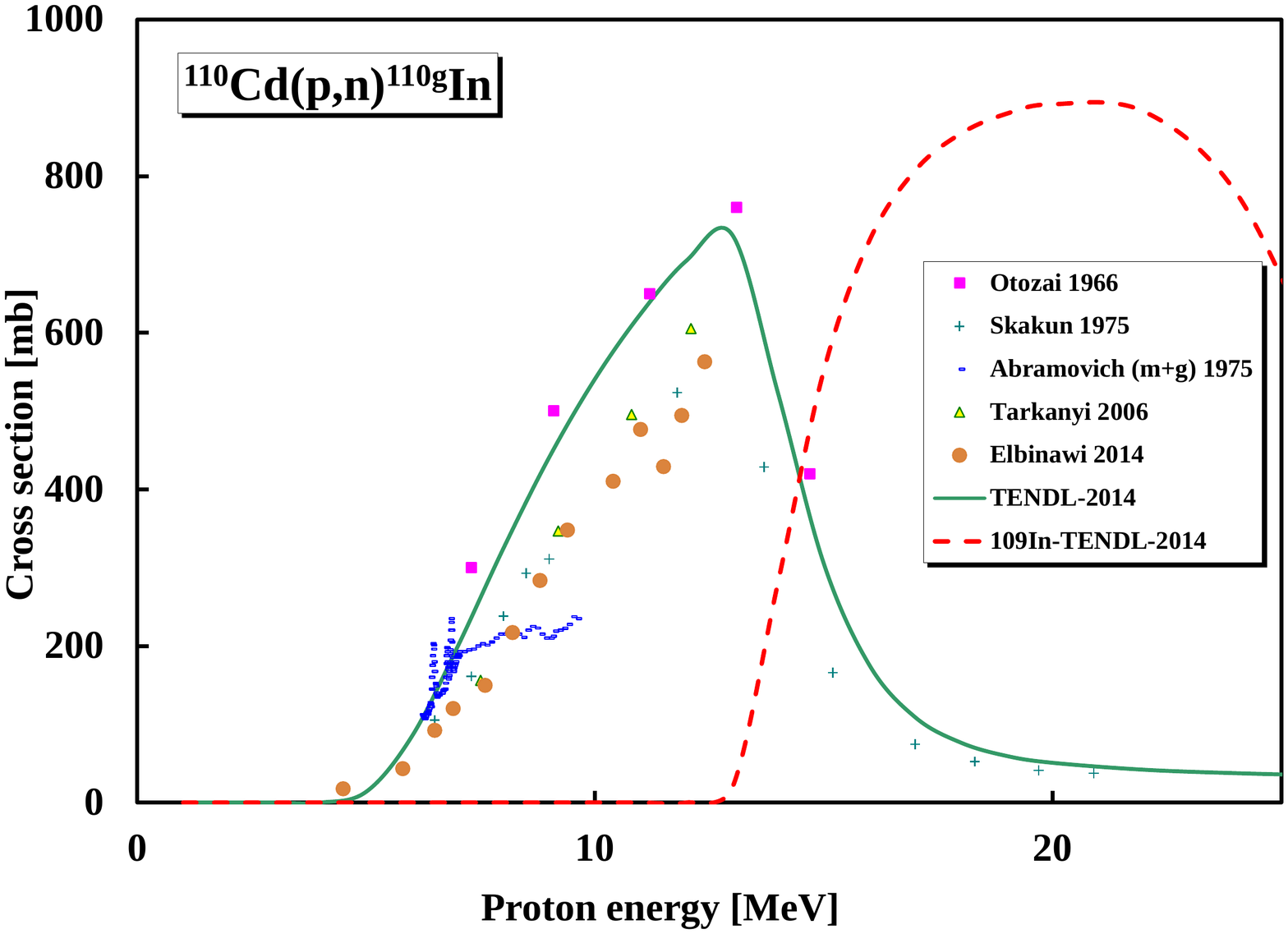}
\caption{Excitation functions of the  $^{110}$Cd(p,n)$^{110m}$In  and $^{110}$Cd(p,2n)$^{109}$In reactions}
\end{figure}

\subsection{Cd+d}
\label{5.2}
As it was mentioned in the previous section on proton induced reactions on $^{nat}$Cd, large amounts of different long-lived radioisotopes of indium are produced when using natural cadmium targets, hence the only realistic candidate is the $^{110}$Cd(d,2n) reaction.
Only a few measurements were published for the $^{110}$Cd(d,2n)$^{110m,110g}$In  reactions: Usher et al. 1977 \cite{36}, Mukhamedov et al. 1983 \cite{37} and Tarkanyi et al. 2007 \cite{4}. The excitation functions are shown in Figs. 8-9. In Fig. 8 the theoretical excitation functions for production of $^{109}$In and $^{111}$In are also shown to illustrate the shape and the magnitude of these simultaneously produced   neighboring indium radioisotopes. According to Figs. 8-9 a very narrow energy window around 15 MeV exists, where the $^{110m}$In yield is high and the co-produced $^{109}$In and $^{111}$In amounts are the lowest. A significant impurity of $^{110g}$In, practically independently from the covered energy range, will however be co-produced as the values of the cross sections are about the same.

\begin{figure}
\includegraphics[scale=0.3]{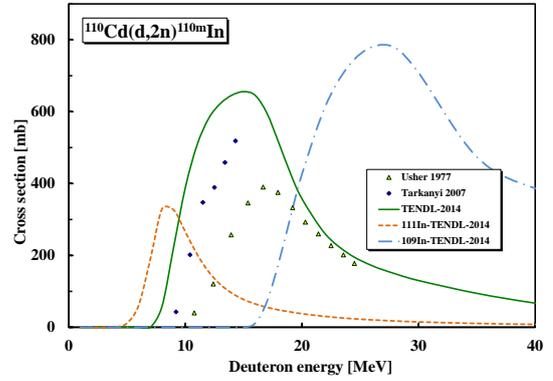}
\caption{Excitation functions of the  $^{110}$Cd(d,xn)$^{109,110m,111}$In reactions}
\end{figure}

\begin{figure}
\includegraphics[scale=0.3]{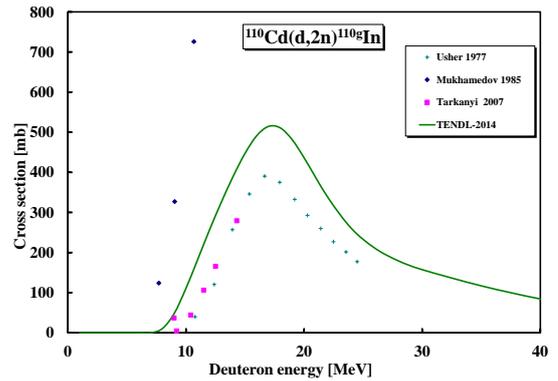}
\caption{Excitation functions of the  $^{110}$Cd(d,2n)$^{110g}$In nuclear reaction}
\end{figure}

\subsection{Ag+$\alpha$}
\label{5.3}
Other possible target – particle combinations for direct production of $^{110m}$In are the Ag+$\alpha$ and Ag+$^{3}$He reactions. Silver has two stable isotopes: $^{107}$Ag (abundance 51.839 \%) and $^{109}$Ag (48.161 \%). When considering alpha induced reactions, the $^{107}$Ag($\alpha$,n) is the main candidate as it allows to minimize co-produced contaminants. Cross sections for direct production of $^{110m,g}$In through alpha induced nuclear reactions on silver have been measured by Misealides et al. \cite{38}, Chaubey et al. \cite{39}, Fukushima et al. \cite{40}, Takacs et al. \cite{7} (our work),  Wasilewsky et al. \cite{41}, Omori et al. \cite{42}, Singh et al. \cite{43} and in the present work. The excitation functions of the $^{107}$Ag($\alpha$,xn)$^{110m}$In, $^{109}$In, $^{110g}$In reactions are shown in Fig 10-12. According to Fig. 12 the production of $^{109}$In is starting only above 16 MeV. The impurity from the ground state $^{110g}$In is high over the whole energy range. 
From literature it is well known that when $^{nat}$Cd targets are used, longer-lived reaction products are present practically in the whole energy range. Especially for the medically important $^{111}$In, most frequently produced though the $^{112}$Cd(p,2n) reaction with 24-25 MeV incident proton energy,  the $^{109}$Ag($\alpha$,2n)  could be a real alternative production route, taking into account that the $^{111}$In can be produced with high yields and low impurity levels.  A large number  of measurements were done for the  $^{109}$Ag($\alpha$,2n)$^{111}$In  reaction: Patel et al. \cite{44} Hasbroek et al. \cite{45}, Ismail et al. \cite{46}, Wasilewsky et al. \cite{47}, Takacs et al. \cite{7}, Mukherjee et al. \cite{48}, Guin et al. \cite{49}, Chaubey \cite{39}, Fukushima \cite{40}, Chuvilskaya \cite{50}, Peng \cite{51}, Bleuer \cite{52},  Porges \cite{53}, Xiufeng \cite{54}, Singh et al. \cite{43} and the present work. The experimental results are shown in Fig. 13.

\begin{figure}
\includegraphics[scale=0.3]{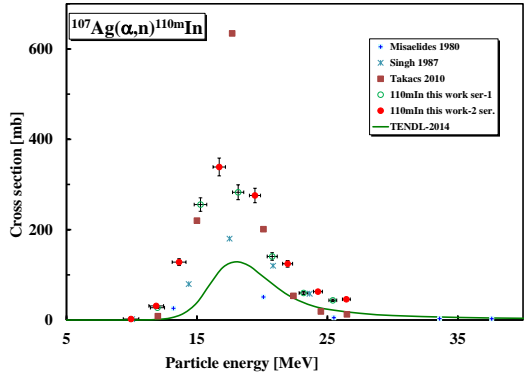}
\caption{Excitation functions of the  $^{107}$Ag($\alpha$,n)$^{110m}$In nuclear reaction}
\end{figure}

\begin{figure}
\includegraphics[scale=0.3]{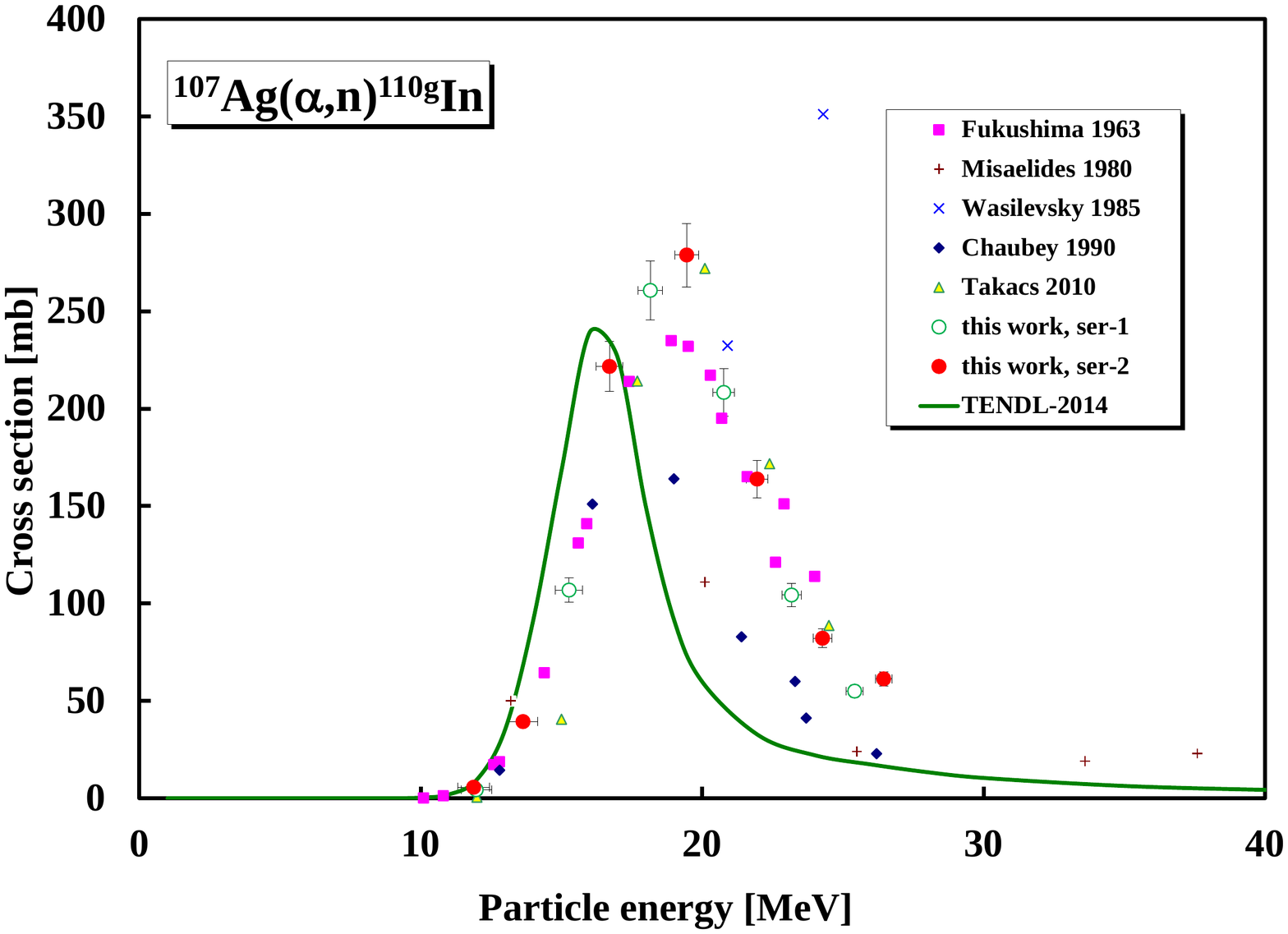}
\caption{Excitation functions of the  $^{107}$Ag($\alpha$,n)$^{110g}$In nuclear reaction}
\end{figure}

\begin{figure}
\includegraphics[scale=0.3]{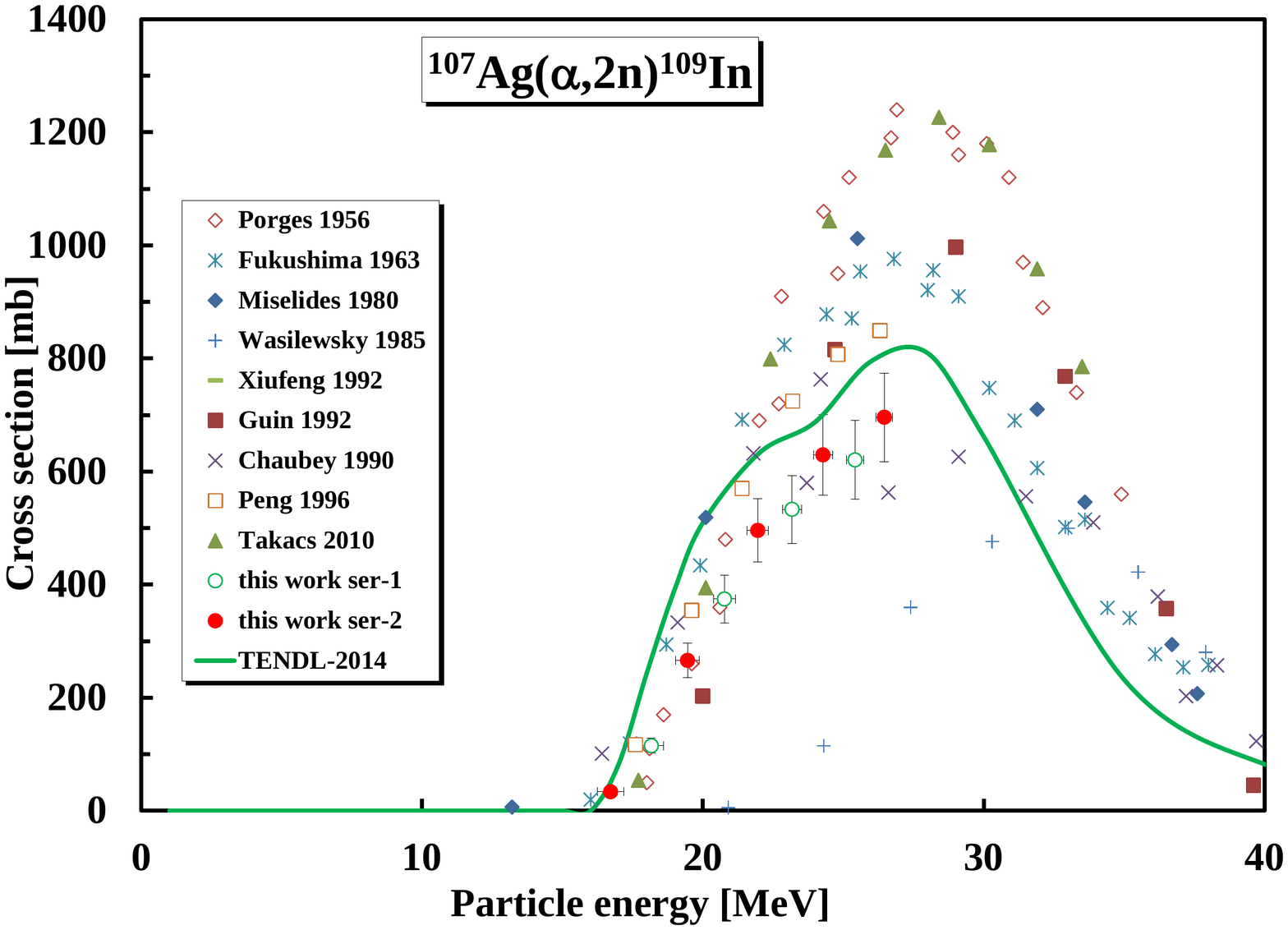}
\caption{Excitation functions of the  $^{107}$Ag($\alpha$,2n)$^{109}$In nuclear reaction}
\end{figure}

\begin{figure}
\includegraphics[scale=0.3]{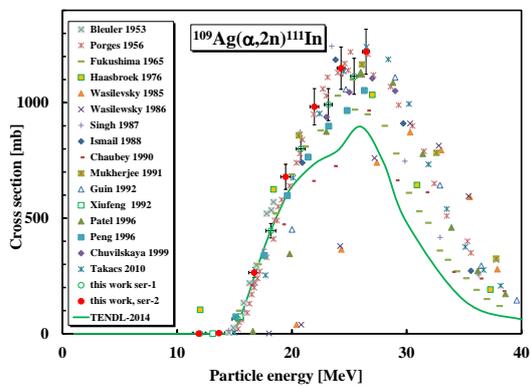}
\caption{Excitation functions of the  $^{109}$Ag($\alpha$,2n)$^{111}$In   nuclear reaction}
\end{figure}

\subsection{Ag+$^3$He}
\label{5.4}
When using $^{3}$He beams the useful reaction on Ag is $^{109}$Ag($^{3}$He,2n). The disturbing products are $^{109}$In, $^{110g}$In and $^{111}$In. The excitation functions for $^{109}$Ag($^{3}$He,xn)$^{109,110m,110g,111}$In are shown in Figs 14-17. The low reliability of predictions of TENDL-2014 for $^{3}$He induced processes can be remarked for these four reactions.
The experimental cross section data on the $^{109}$Ag($^{3}$He,2n)110m,gIn reactions were measured by Misaelides et al. \cite{38}, Marten et al. \cite{55}, Nagame  et al. \cite{56}, Omori et al. \cite{57} and in this work.
According to Fig. 14, the cross sections for production of $^{110m}$In are low. The reaction can effectively be used in the 10-30 MeV range but over the whole energy range a significant yield for $^{109}$In is seen (Fig. 17). Small cross sections for $^{111}$In production are also present over the whole energy range (Fig. 16) and for $^{110g}$In the cross sections are similar to those for $^{110m}$In (Fig. 15).  

\begin{figure}
\includegraphics[scale=0.3]{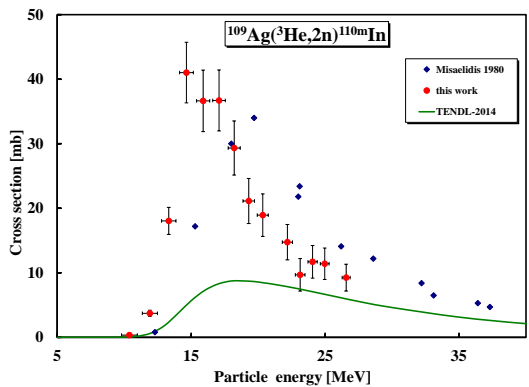}
\caption{Excitation functions of the  $^{109}$Ag($^{3}$He,xn)$^{110m}$In nuclear reaction}
\end{figure}

\begin{figure}
\includegraphics[scale=0.3]{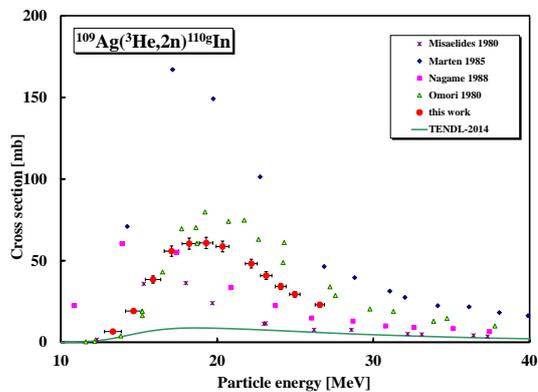}
\caption{Excitation functions of the  $^{109}$Ag($^{3}$He,2n)$^{110g}$In   nuclear reaction}
\end{figure}

\begin{figure}
\includegraphics[scale=0.3]{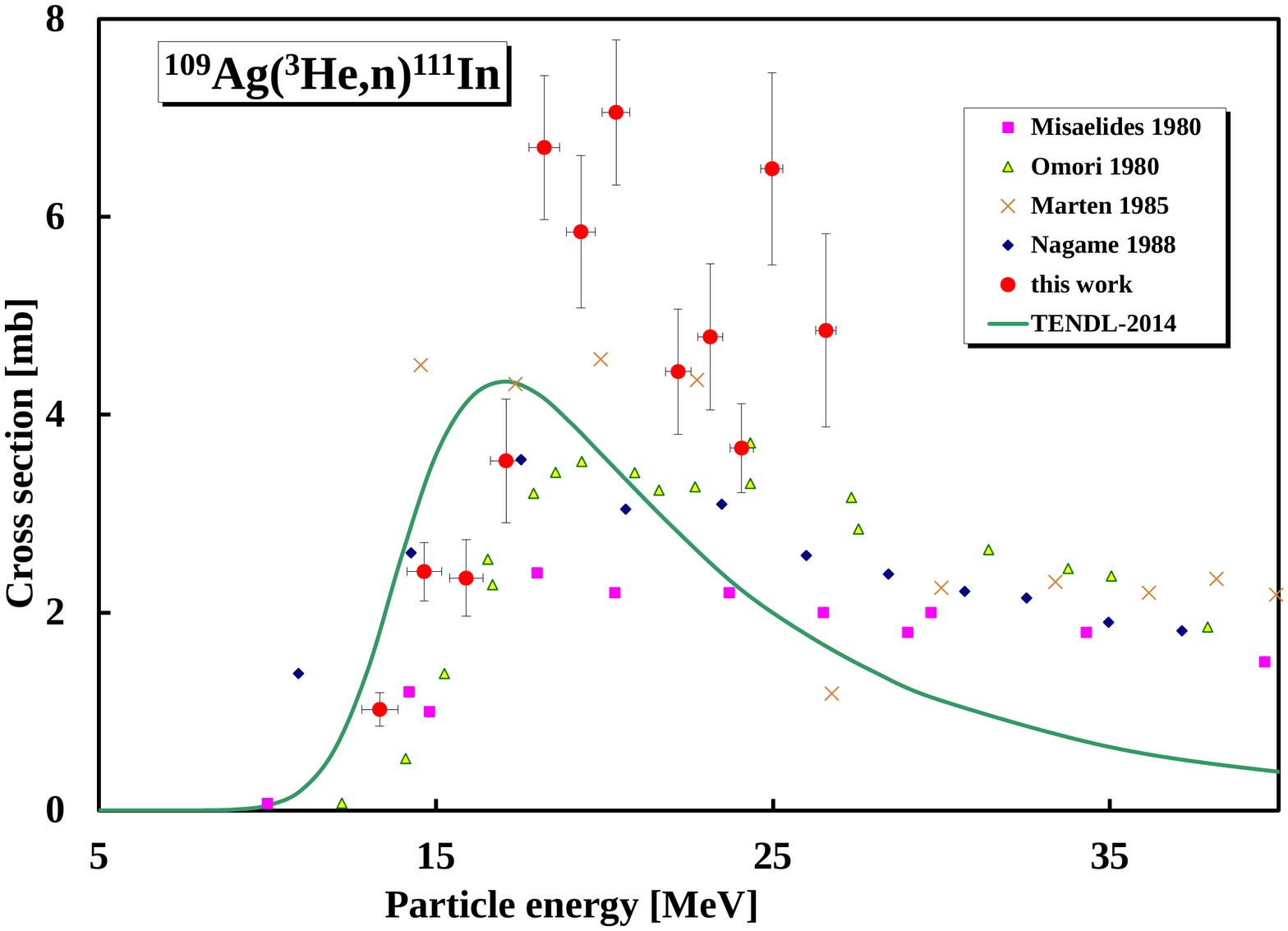}
\caption{Excitation functions of the   $^{109}$Ag($^{3}$He,n) $^{111}$In nuclear reaction}
\end{figure}

\begin{figure}
\includegraphics[scale=0.3]{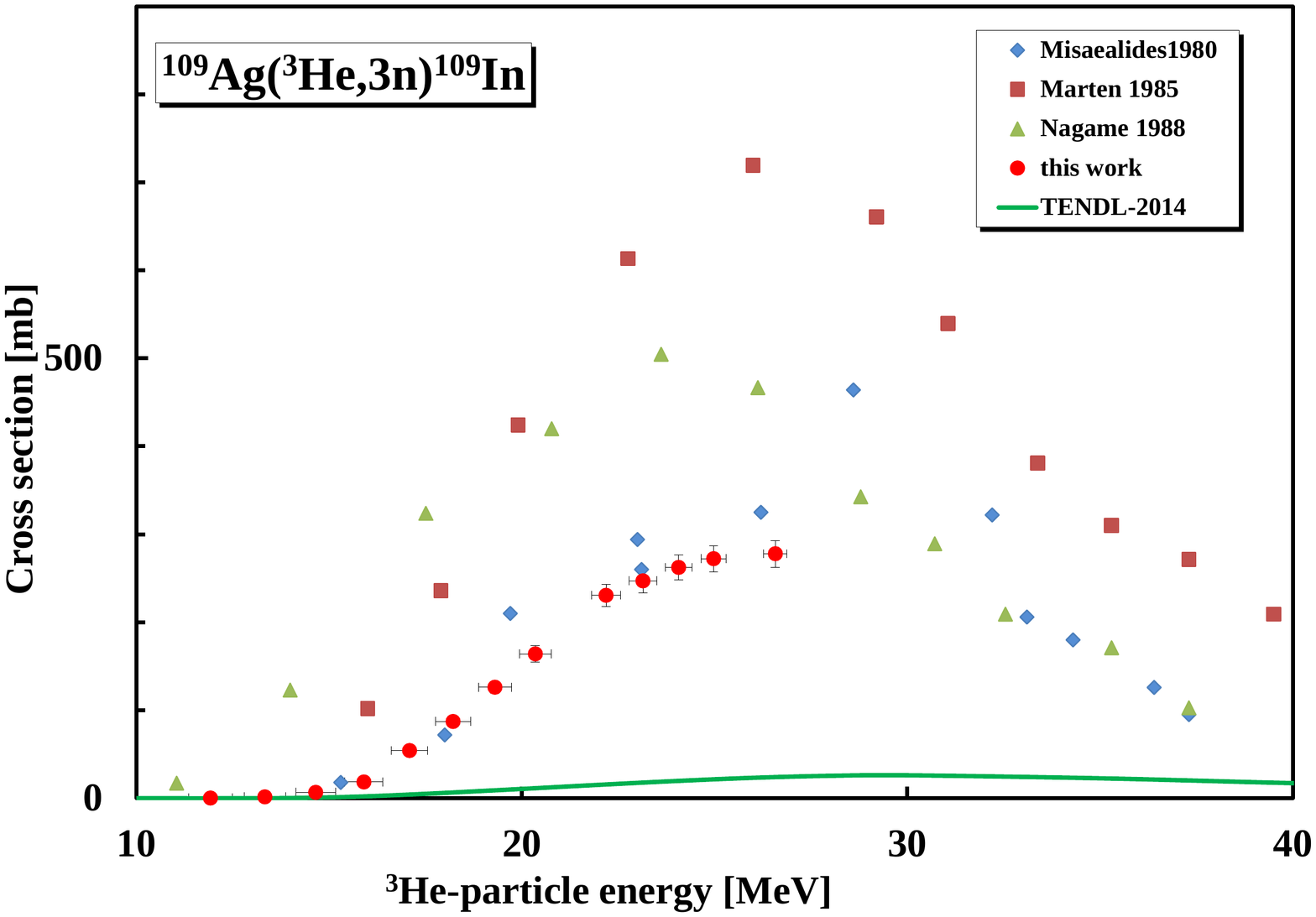}
\caption{Excitation functions of the  $^{109}$Ag($^{3}$He,3n)$^{109}$In nuclear reaction}
\end{figure}

\begin{table*}[t]
\tiny
\caption{Activation cross sections of the $^{nat}$In(p,xn)$^{110}$Sn reaction}
\centering
\begin{center}
\begin{tabular}{|r|r|r|r|r|r|r|r|}
\hline
\multicolumn{4}{|c|}{\textbf{CYRIC}} & \multicolumn{4}{|c|}{\textbf{LLN}} \\
\hline
\textbf{E (MeV)} & \textbf{$\Delta$E (MeV} & \textbf{$\sigma$ (mbarn)} & \textbf{$\Delta\sigma$ (mbarn)} & \textbf{E (MeV)} & \textbf{$\Delta$E (MeV} & \textbf{$\sigma$ (mbarn)} & \textbf{$\Delta\sigma$ (mbarn)} \\
\hline
59.5 & 0.5 & 17.0 & 1.9 & 37.5 & 0.8 & 3.7 & 0.4 \\
\hline
60.3 & 0.5 & 23.2 & 2.6 & 39.5 & 0.8 & 7.1 & 0.8 \\
\hline
61.2 & 0.4 & 29.2 & 3.3 & 41.3 & 0.7 & 10.0 & 1.1 \\
\hline
62.1 & 0.4 & 34.3 & 3.9 & 43.1 & 0.7 & 12.0 & 1.3 \\
\hline
62.9 & 0.4 & 50.6 & 5.7 & 44.8 & 0.7 & 13.0 & 1.5 \\
\hline
63.8 & 0.4 & 64.1 & 7.2 & 46.5 & 0.6 & 13.5 & 1.5 \\
\hline
64.6 & 0.4 & 63.8 & 7.2 & 48.1 & 0.6 & 13.0 & 1.5 \\
\hline
65.5 & 0.4 & 71.6 & 8.1 & 49.7 & 0.6 & 12.0 & 1.4 \\
\hline
66.3 & 0.4 & 79.5 & 9.0 & 51.2 & 0.5 & 11.0 & 1.2 \\
\hline
67.1 & 0.3 & 84.3 & 9.5 & 52.7 & 0.5 & 10.2 & 1.1 \\
\hline
67.9 & 0.3 & 88.2 & 9.9 & 54.2 & 0.5 & 10.3 & 1.2 \\
\hline
68.7 & 0.3 & 89.5 & 10.1 & 55.7 & 0.5 & 12.4 & 1.4 \\
\hline
69.5 & 0.3 & 100.1 & 11.3 & 57.1 & 0.4 & 16.6 & 1.9 \\
\hline
 & & & & 58.4 & 0.4 & 23.2 & 2.6 \\
\hline
 & & & & 59.8 & 0.4 & 32.3 & 3.6 \\
\hline
 & & & & 61.1 & 0.4 & 42.4 & 4.8 \\
\hline
 & & & & 62.4 & 0.3 & 53.0 & 6.0 \\
\hline
 & & & & 63.7 & 0.3 & 62.3 & 7.0 \\
\hline
 & & & & 65.0 & 0.3 & 70.9 & 8.0 \\
\hline
\end{tabular}

\end{center}
\end{table*} 

\begin{table*}[t]
\tiny
\caption{Activation cross sections of the natAg($^{3}$He,xn)$^{109,110,,110g,111}$In reactions}
\centering
\begin{center}
\begin{tabular}{|r|r|r|r|r|r|r|r|r|r|}
\hline
\multicolumn{2}{|c|}{\textbf{Energy}} & \multicolumn{2}{|c|}{$^{109}$In} & \multicolumn{2}{|c|}{$^{110m}$In} & \multicolumn{2}{|c|}{$^{110g}$In} & 
\multicolumn{2}{|c|}{$^{111}$In} \\
\hline
\multicolumn{2}{|c|}{\textbf{E $\pm \Delta$\newline E(MeV)}} & \multicolumn{8}{|c|}{\textbf{$\sigma\pm\Delta\sigma$\newline (mbarn)}} \\
\hline
10.4 & 0.6 & & & 0.3 & 0.1 & & & &  \\
\hline
11.9 & 0.6 & 0.4 & 0.1 & 3.7 & 0.5 & & & &  \\
\hline
13.3 & 0.5 & 1.6 & 0.2 & 18.0 & 2.1 & 0.5 & 6.6 & 1.0 & 0.4 \\
\hline
14.7 & 0.5 & 6.8 & 0.6 & 41.0 & 4.7 & 1.2 & 19.1 & 2.4 & 0.6 \\
\hline
15.9 & 0.5 & 18.4 & 1.5 & 36.6 & 4.8 & 2.2 & 38.4 & 2.3 & 0.8 \\
\hline
17.1 & 0.5 & 54.2 & 3.2 & 36.7 & 4.7 & 3.2 & 55.8 & 3.5 & 1.3 \\
\hline
18.2 & 0.5 & 87.4 & 5.0 & 29.3 & 4.2 & 3.4 & 60.4 & 6.7 & 1.5 \\
\hline
19.3 & 0.4 & 126.1 & 7.0 & 21.1 & 3.5 & 3.5 & 60.8 & 5.8 & 1.6 \\
\hline
20.3 & 0.4 & 164.2 & 9.1 & 18.9 & 3.3 & 3.3 & 58.7 & 7.1 & 1.5 \\
\hline
22.2 & 0.4 & 230.7 & 12.6 & 14.7 & 2.7 & 2.7 & 48.2 & 4.4 & 1.3 \\
\hline
23.1 & 0.4 & 247.1 & 13.5 & 9.7 & 2.5 & 2.3 & 40.8 & 4.8 & 1.5 \\
\hline
24.1 & 0.3 & 262.4 & 14.3 & 11.7 & 2.5 & 1.9 & 34.2 & 3.7 & 0.9 \\
\hline
25.0 & 0.3 & 272.2 & 14.9 & 11.4 & 2.4 & 1.7 & 29.4 & 6.5 & 2.0 \\
\hline
26.6 & 0.3 & 277.8 & 15.1 & 9.3 & 2.1 & 1.4 & 23.0 & 4.9 & 2.0 \\
\hline
\end{tabular}

\end{center}
\end{table*} 

\begin{table*}[t]
\tiny
\caption{Activation cross sections of the natAg($\alpha$,xn)$^{109,110,,110g,111}$In reactions}
\centering
\begin{center}
\begin{tabular}{|r|r|r|r|r|r|r|r|r|r|}
\hline
\multicolumn{2}{|c|}{\textbf{Energy}} & \multicolumn{2}{|c|}{$^{109}$In} & \multicolumn{2}{|c|}{$^{110m}$In} & \multicolumn{2}{|c|}{$^{110g}$In} & 
\multicolumn{2}{|c|}{$^{111}$In} \\
\hline
\multicolumn{2}{|c|}{\textbf{E $\pm \Delta$\newline E(MeV)}} & \multicolumn{8}{|c|}{\textbf{$\sigma\pm\Delta\sigma$\newline (mbarn)}} \\
\hline
9.9 & 0.6 & & & 2.3 & 0.1 & & & & \\
\hline
11.9 & 0.6 & & & 31.2 & 1.8 & 5.6 & 0.4 & 0.3 & 0.02 \\
\hline
12.0 & 0.5 & & & 27.2 & 1.6 & 4.4 & 0.3 & 0.3 & 0.03 \\
\hline
13.6 & 0.5 & & & 128.0 & 7.5 & 39.3 & 2.3 & 2.5 & 0.14 \\
\hline
15.3 & 0.5 & & & 255.6 & 14.9 & 106.8 & 6.2 & 63.3 & 3.4 \\
\hline
16.7 & 0.5 & 33.8 & 4.7 & 338.8 & 19.7 & 221.7 & 12.9 & 264.6 & 14.3 \\
\hline
18.2 & 0.4 & 115.0 & 13.2 & 282.8 & 16.5 & 260.7 & 15.2 & 445.7 & 24.1 
\\
\hline
19.5 & 0.4 & 266.0 & 30.2 & 275.6 & 16.0 & 278.8 & 16.3 & 679.1 & 36.7 
\\
\hline
20.8 & 0.4 & 374.5 & 42.2 & 140.7 & 8.2 & 208.3 & 12.2 & 800.2 & 43.3 \\
\hline
22.0 & 0.4 & 495.8 & 56.0 & 124.5 & 7.3 & 163.9 & 9.6 & 982.5 & 53.1 \\
\hline
23.2 & 0.3 & 533.2 & 60.0 & 59.8 & 3.5 & 104.3 & 6.1 & 991.9 & 53.6 \\
\hline
24.3 & 0.3 & 629.4 & 70.9 & 62.8 & 3.7 & 82.1 & 4.8 & 1149.5 & 62.1 \\
\hline
25.4 & 0.3 & 620.6 & 69.8 & 43.6 & 2.5 & 55.0 & 3.2 & 1114.2 & 60.2 \\
\hline
26.5 & 0.3 & 695.9 & 78.2 & 45.9 & 2.7 & 61.3 & 3.6 & 1220.6 & 66.0 \\
\hline
\end{tabular}

\end{center}
\end{table*} 

\section{Integral yields}
\label{6}

Integral yields as a function of energy were calculated by using fitted experimental and/or theoretical cross sections for production of $^{110}$Sn via the $^{113}$In(p,4n), $^{nat}$In(p,xn), $^{108}$Cd($\alpha$,2n), $^{nat}$Cd($\alpha$,xn), $^{110}$Cd($^{3}$He,3n) and $^{nat}$Cd($^{3}$He,xn) reactions and for direct production of $^{110m}$In via $^{110}$Cd(p,n), $^{110}$Cd(d,2n), $^{107}$Ag($\alpha$,xn) and $^{109}$Ag($^{3}$He,2n) reactions (Figs. 18-19). There are only a few experimental thick target yields on these target-reaction combinations measured by Dmitriev \cite{58, 59}, Nickles et al. \cite{60}, Mukhamedov \cite{37} and Abe et al. \cite{61}. Where an energy overlap is existing, our data were compared with those literature values.

\begin{figure}
\includegraphics[scale=0.3]{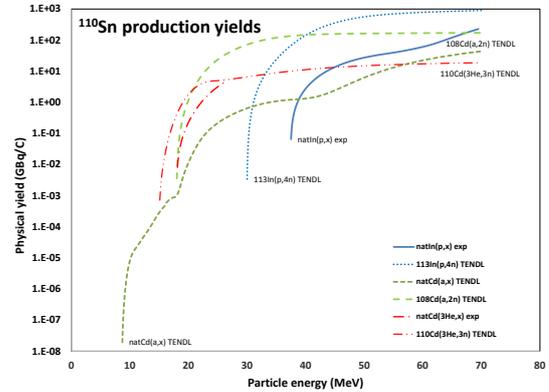}
\caption{Thick target yields for production of $^{110}$Sn}
\end{figure}

\begin{figure}
\includegraphics[scale=0.3]{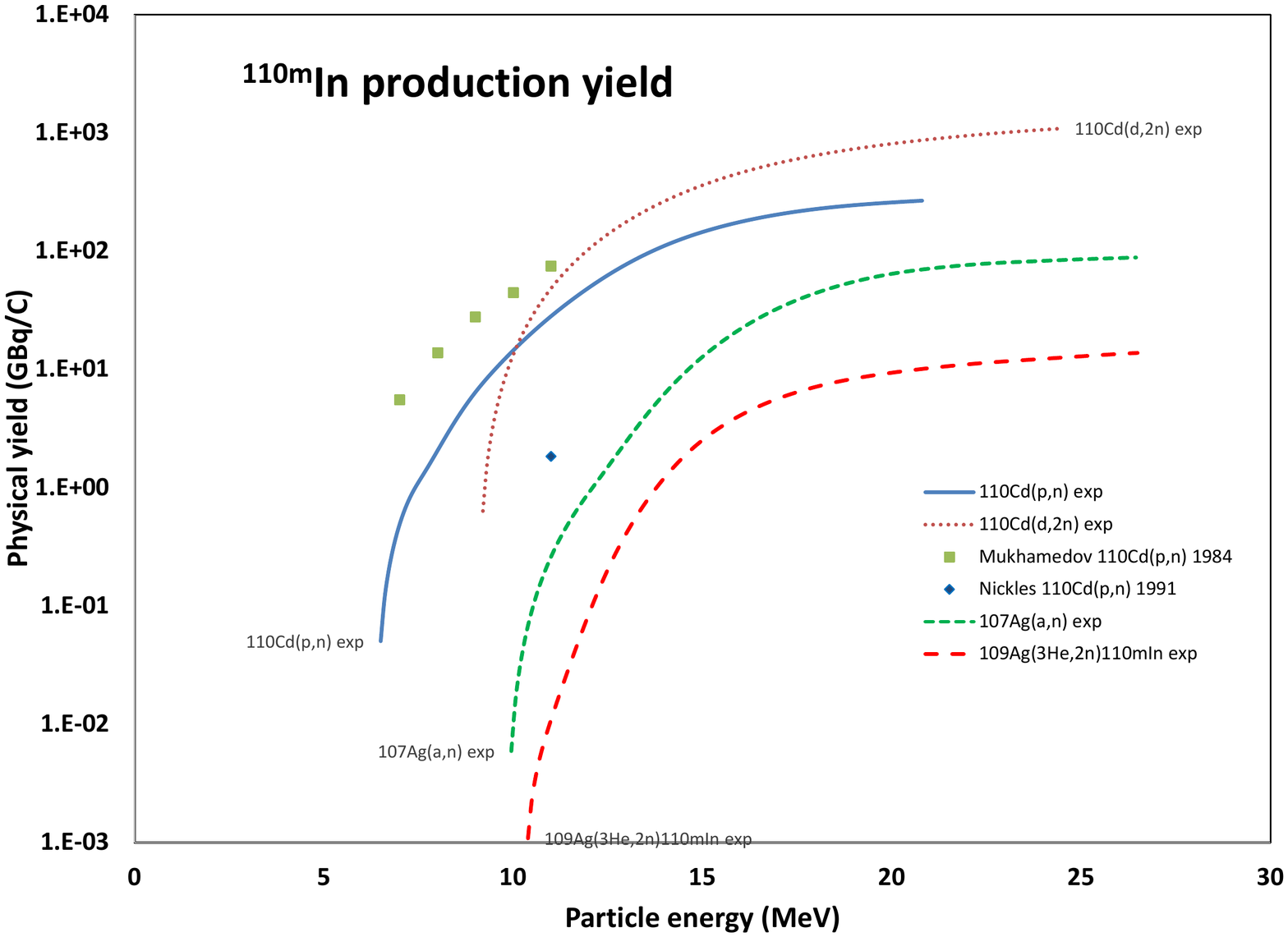}
\caption{Thick target yields for production of $^{110}$mIn}
\end{figure}

\section{Summary}
\label{7}

In the frame of a systematic study of production routes of the medically useful $^{110m}$In, experimental cross section data for the $^{nat}$In(p,xn)$^{110}$Sn indirect route and for $^{nat}$Ag($\alpha$,xn)$^{109,110m,110g,111}$In and natAg($^{3}$He,xn)$^{109,110m,110g,111}$In direct nuclear reactions were measured.
The new data are in good agreement with earlier results and are in acceptable agreement with the theoretical predictions in TENDL-2014 except for the $^{3}$He induced reactions. Thick target yields were derived for different routes relevant for production of the radioisotope of interest $^{110m}$In. The $^{110}$Sn($^{110m}$In) generator could be prepared at low and medium energies with light charged particles via the $^{113}$In(p,4n), $^{nat}$In(p,xn), $^{110}$Cd($\alpha$,2n), $^{nat}$Cd($\alpha$,xn), $^{110}$Cd($^{3}$He,3n) and $^{nat}$Cd($^{3}$He,xn) reactions. Each of these routes requires high incident energy and a proper selection of an adapted energy range. The main advantage of the indirect method is the high radionuclide purity, which can easily be assured by the proper irradiation and cooling parameters. Another advantage is that, depending on the required specific activity, natural targets can be used. Considering the available commercial accelerators, use of the $^{nat}$In(p,xn)$^{110}$Sn reaction seems to be the simplest and most productive method, however requiring 70-100 MeV beam energy. It should be mentioned that the generator can be produced also at lower energy machines by using 30 MeV alpha beams and enriched $^{108}$Cd targets.
For direct production the $^{110}$Cd(p,n), $^{110}$Cd(d,2n), $^{107}$Ag($\alpha$,xn) and $^{109}$Ag($^{3}$He,2n) reactions are the most suitable candidate routes. Lower energy accelerators can also be used, but highly enriched targets are required. Although the radionuclide impurity level is lower, significant amount of $^{110g}$In are always present in the product. If the $^{110g}$In level is not taken into account, the $^{110}$Cd(p,n) reaction seems the most promising production route.

\section{Acknowledgements}

This work was performed in the frame of the HAS-FWO Vlaanderen (Hungary-Belgium) project. The authors acknowledge the support of the research project and of the respective institutions (CYRIC, VUB, LLN) in providing the beam time and experimental facilities.
%\FloatBarrier
 


%% References

%\clearpage
\bibliographystyle{elsarticle-num}
\bibliography{In110}






% \end{thebibliography}

\end{document}